\documentstyle[12pt,aps,epsf,amssymb]{revtex}
\begin{document}

\newcommand{\ini}{\begin{equation}}
\newcommand{\fin}{\end{equation}}
\newcommand{\inir}{\begin{eqnarray}}
\newcommand{\finr}{\end{eqnarray}}
\newcommand{\inif}{\begin{figure}}
\newcommand{\finf}{\end{figure}}
\newcommand{\bc}{\begin{center}}
\newcommand{\ec}{\end{center}}
\def\ol{\overline}
\def\pa{\partial}
\def\ra{\rightarrow}
\def\ts{\times}
\def\df{\dotfill}
\def\bs{\backslash}

\hfill DSF-T-99/14

\vspace{1cm}

\centerline{\LARGE{Quark mass matrices}}

\centerline{\LARGE{and observable quantities}}

\vspace{1 cm}

\centerline{\large{D. Falcone}}

\vspace{1 cm}

\centerline{Dipartimento di Scienze Fisiche, Universit\`a di Napoli,}
\centerline{Mostra d'Oltremare, Pad. 19, I-80125, Napoli, Italy.}

\centerline{{\tt e-mail:falcone@na.infn.it}}

\vspace{1 cm}

\begin{abstract}

\centerline{ABSTRACT}

\noindent
A nearly historical account of quark mass matrix models is given,
and the structure of quark mass matrices in the Standard Model is studied.
For a minimal parameter basis suggested earlier, where $M_u$ is diagonal and
$M_{d11}$, $M_{d13}$, $M_{d31}$ are zero, the dependence
of mass matrices on the CP violating phase $\delta$ of $V_{CKM}$ is
reported: all parameters are almost independent, except $M_{d22}$ and
$M_{d23}$, and the equality $|M_{d22}|=M_{d23}$ is obtained for a value of
$\delta$ very close to the value which is favoured by experiments.
Moreover, on this basis, $M_{d12} \simeq M_{d21}$ and
$M_{d33} \simeq 2M_{d32}$. Some comments on mass matrices in left-right
symmetric models  are added.

$~$

\noindent
PACS numbers: 12.15.Ff, 12.15.Hh

\end{abstract}

\newpage

\section{ Introduction}

\noindent
Understanding the pattern of fermion masses and mixings is a key problem of
particle physics. In this paper we study quark masses and mixings in the
framework of the standard $SU(3) \times SU(2) \times U(1)$ gauge theory.
We discuss some possible forms of quark mass matrices at the scale $\mu=M_Z$,
and what they suggest. It is well known that the Standard Model (SM)
does not predict fermion parameters, and looking at the structure of mass
matrices we can yield hints towards a more fundamental theory.

The part of the SM Lagrangian we are interested in is formed
by the quark mass and charged weak current terms
\ini
L_M=\ol{u}_L M_u u_R + \ol{d}_L M_d d_R + g \ol{u}_L d_L W.
\fin
When we diagonalize the mass matrices $M_u$ and $M_d$ we get (renaming the
quark fields)
\ini
L_D=\ol{u}_L D_u u_R + \ol{d}_L D_d d_R + g \ol{u}_L  V_{CKM} d_L W
\fin
where the mixing matrix $V_{CKM}$ \cite{ckm} contains three angles
and one phase.
For the standard and Wolfenstein \cite{wol} parameterizations of $V_{CKM}$
we refer to \cite{pdg}.
For two generations the mixing matrix becomes a rotation
\ini
V_C=\left( \begin{array}{cc}
               \cos \theta_C & \sin \theta_C \\
               -\sin \theta_C & \cos \theta_C
              \end{array}\right)
\fin
where $\theta_C$ is the Cabibbo angle. We take the numerical values of
quark masses
(at $\mu=M_Z$) from ref.\cite{fk} and of the mixing angles from ref.\cite{pdg};
we set $\lambda=0.22$, with $\sin \theta_C \simeq \lambda$.
There is a clear hierarchy of quark masses: $m_u \ll m_c \ll m_t$ and
$m_d \ll m_s \ll m_b$; $m_b \ll m_t$. Moreover, $V_{CKM}$ is near the identity and
$V_{ub} \ll V_{cb} \ll V_{us}$.

$L_D$ is expressed by means of the ten observable quantities. On the contrary,
$M_u$ and $M_d$, appearing in $L_M$, contain, in general,
eighteen real parameters each.
As a matter of fact, in the SM, it is possible
to perform, without physical consequences, the following unitary
transformations on the quark fields
\ini
u_L \ra U u_L,~d_L \ra U d_L;
\fin
\ini
u_R \ra V_u u_R,~d_R \ra V_d d_R.
\fin                                       
In particular we can absorb some non physical phases in $M_u$ and $M_d$
by means of the transformations
\ini
u_L \ra $diag$(1,e^{i \varphi_1},e^{i \varphi_2})u_L,~
d_L \ra $diag$(1,e^{i \varphi_1},e^{i \varphi_2})d_L;
\fin
\ini
u_R \ra $diag$(e^{i \varphi_3},e^{i \varphi_4},e^{i \varphi_5})u_R,~
d_R \ra $diag$(e^{i \varphi_6},e^{i \varphi_7},e^{i \varphi_8})d_R.
\fin
In this way a maximum of five phases in $M_1$ and of three in $M_2$
can be absorbed.
Using transformations (4),(5) (and (6),(7)) one can also get zeros in mass
matrices or
relations between elements, and reduce the number of independent parameters
to ten. When such a basis is achieved we talk about a minimal parameter (m.p.)
basis. For example, one can yield hermitian mass matrices \cite{fj}.
In fact, by a polar decomposition theorem,
$M=HX$, where $H$ is hermitian and $X$ is unitary, and by means of
different $V_u$, $V_d$, hermitian mass matrices can be obtained.
In particular, it is always possible to choose one matrix diagonal and the other
hermitian \cite{ma}. In fact, $M_1$ can be diagonalized by a biunitary transformation
$U^+ M_1 V_1$, and $M_2$ becomes hermitian by the product $U^+ M_2 V_2$.
With hermitian matrices one can then take $U=V_u=V_d$.
Nevertheless the freedom in eqns.(4),(5) was not much used to get zeros
till the paper \cite{blm}.
Hence, our paper is divided in two main parts: in section II we review hermitian
quark
mass matrix models and in section III we study
mass matrices on some interesting weak bases. One of such bases \cite{ft}
leads to a predictive reduction of independent parameters.
In the final section we match
the two approaches and try some conclusions in the context of a left-right
symmetric gauge group.

\section{ Review}

\noindent
The starting point of quark mass matrix models can be considered to rise from
the following relation between the Cabibbo angle and a meson mass ratio
\cite{gsto}:
\ini
\sin^2 \theta_C \simeq \frac {m_{\pi}^2}{ 2 m_{K}^2}.
\fin
In fact, the r.h.s. of eqn.(8) is related to a quark mass ratio \cite{w}, 
\ini
\frac {m_{\pi}^2}{ 2 m_{K}^2} \simeq \frac {m_d}{m_s},
\fin
and then
\ini
\sin \theta_C \simeq \sqrt{\frac {m_d}{m_s}}.
\fin
Weinberg \cite{w} obtained such a relation from a particular structure of quark
mass matrices. Let us consider a real symmetric matrix
\ini
M=\left( \begin{array}{cc}
          A & B \\
          B & D
          \end {array}\right)
\fin
which can be diagonalized by an orthogonal transformation
$O^T M O = D$ with a mixing angle $\theta$ given by
\ini
\tan 2 \theta= \frac{2B}{D-A}.
\fin
In the limit $A \ra 0$ and if $|m_1| \ll |m_2|$ we have also the approximate
eigenvalues
\ini
m_2 \simeq D,~~m_1 \simeq -\frac{B^2}{D}
\fin
and the relation
\ini
\sin \theta \simeq \frac{B}{D} \simeq \sqrt{-\frac{m_1}{m_2}}.
\fin
Then we have $B \simeq \sqrt{-m_1 m_2}$ and
\ini
M \simeq \left( \begin{array}{cc}
          0 & \sqrt{-m_1 m_2} \\
          \sqrt{-m_1 m_2} & m_2
          \end {array}\right).
\fin     
Now, $m_1$ (or $m_2$) is negative, and if $M=M_d$ we obtain
\ini
D_d=\left( \begin{array}{cc}
          -m_d & 0 \\
          0 & m_s
          \end {array}\right),
\fin            
\ini
\sin \theta \simeq \sqrt{\frac{m_d}{m_s}}.
\fin
Actually, in Weinberg's paper, starting from an arbitrary real matrix
\ini
M=\left( \begin{array}{cc}
          A & B \\
          C & D
          \end {array}\right),
\fin
one first gets $A=0$ by an orthogonal transformation on right-handed fields,
and then assumes $B=C$.
In any case, for the first two generations of quarks, if we set $M_u=D_u$ and
\ini
M_d=\left( \begin{array}{cc}
          0 & B \\
          B & D
          \end {array}\right),
\fin
we have $B \ll D$ and formula (10).            
We can now express the quark mass matrices in terms of the quark masses by
\ini
M_d \simeq \left( \begin{array}{cc}
          0 & \sqrt{m_d m_s} \\
          \sqrt{m_d m_s} & m_s
          \end {array}\right),~
     M_u=\left( \begin{array}{cc}
          m_u & 0 \\
          0 & m_c
          \end {array}\right).
\fin            
In view of the fact that the elementary particle theory is probably more
symmetric than the SM, we are led to consider a similar structure of both
mass matrices.
This was achieved in the famous Fritzsch model \cite{f}.
For two generations 
\ini
M_d=\left( \begin{array}{cc}
          0 & B \\
          B & D
          \end {array}\right),~M_u=\left( \begin{array}{cc}
                                           0 & B' \\
                                           B' & D'
                                       \end {array}\right)
\fin
imply $B \ll D$, $B' \ll D'$ and
\ini
\sin \theta_C \simeq \sqrt{\frac {m_d}{m_s}}-\sqrt{\frac {m_u}{m_c}}
\approx \sqrt{\frac {m_d}{m_s}}.
\fin
The first generation gets its mass by mixing with the heavier second generation
and we have
\ini
M_d \simeq \left( \begin{array}{cc}
          0 & \sqrt{m_d m_s} \\
          \sqrt{m_d m_s} & m_s
          \end {array}\right),~
    M_u \simeq \left( \begin{array}{cc}
          0 & \sqrt{m_u m_c} \\
          \sqrt{m_u m_c} & m_c
          \end {array}\right).
\fin
Here one has to note that
\ini
M_d \simeq m_s \left( \begin{array}{cc}
          0 & \lambda \\
          \lambda & 1
          \end {array}\right),~M_u \simeq m_c \left( \begin{array}{cc}
                                           0 & \lambda^2 \\
                                           \lambda^2 & 1
                                       \end {array}\right),
\fin
with the zero in position 1-1 which could be approximate: naturalness
\cite{fh,pw} leads to
$M_{d11} \lesssim \lambda^2$, $M_{u11} \lesssim \lambda^4$, respectively.
The hierarchical form of mass matrices in eqn.(24) is due to the hierarchy of quark
masses, that is matrices in eqn.(24) lead to large quark mass ratios and also to
small mixing. It is then important to understand how such a form can arise.
In ref.\cite{fn} a qualitative answer to this question is given by means of
a broken continuous abelian symmetry beyond the SM.

For three generations of quarks the Fritzsch model is given by
\ini
M_d=\left( \begin{array}{ccc}
          0 & A & 0 \\
          A & 0 & B \\
          0 & B & C
          \end {array}\right),
~M_u=\left( \begin{array}{ccc}
          0 & A' & 0 \\
          A'^* & 0 & B' \\
          0 & B'^* & C'
          \end {array}\right),
\fin
and with $A \ll B \ll C$, $A' \ll B' \ll C'$ one yields the relations
\ini
V_{us} \simeq \left|\sqrt{\frac{m_d}{m_s}}-
e^{i\sigma}\sqrt{\frac{m_u}{m_c}}\right|,
\fin
\ini
V_{cb} \simeq \left|\sqrt{\frac{m_s}{m_b}}-
e^{i\rho}\sqrt{\frac{m_c}{m_t}}\right|,
\fin
but setting $m_t=180$ GeV we obtain a too large $V_{cb}$ (from 0.10 to 0.25,
using the central values of quark masses). The value of $\sigma$ must be close
to $\pm \pi/2$, because eqn.(10) is already well satisfied.
In this model the two lighter generations get mass by direct and indirect
mixing with the heavier third generation.
An alternative model, with a different structure of $M_d$,
is due to Georgi and Jarlskog \cite{gj} (but see also ref.\cite{hh}),
\ini
M_d=\left( \begin{array}{ccc}
          0 & A & 0 \\
          A^* & B & 0 \\
          0 & 0 & C
          \end {array}\right),
~M_u=\left( \begin{array}{ccc}
          0 & A' & 0 \\
          A' & 0 & B' \\
          0 & B' & C'
          \end {array}\right),
\fin
which gives a relation similar to (26)
and the new relation
\ini
V_{cb} \simeq \sqrt{\frac{m_c}{m_t}}.
\fin
Thus, $V_{cb}= 0.061 \pm 0.005$ is of the right order but again too large.
We can write
\ini
M_d \simeq \left( \begin{array}{ccc}
          0 & \sqrt{m_d m_s} & 0 \\
          \sqrt{m_d m_s} & m_s & 0 \\
          0 & 0 & m_b
          \end {array}\right),
~M_u \simeq \left( \begin{array}{ccc}
          0 & \sqrt{m_u m_c} & 0 \\
          \sqrt{m_u m_c} & 0 & \sqrt{m_c m_t} \\
          0 & \sqrt{m_c m_t} & m_t
          \end {array}\right).
\fin
Actually, this model has been studied in SU(5) \cite{gj} and
SO(10) \cite{hrr} and also in
supersymmetric versions \cite{dhr}. There the charged lepton mass matrix is
related to $M_d$ and,
in the SO(10) model, one also has predictions on the neutrino sector.

The Georgi-Jarlskog model can be seen as a Fritzsch model for two generations
for $M_d$ plus a Fritzsch model for three generations for $M_u$. Other 
modifications of the Fritzsch model consist in taking elements 1-3 or 2-2
different from zero. Neither the modification of the element 1-3 \cite{als}
nor the element 2-2 in $M_u$ \cite{gujo}, can give the correct $V_{cb}$.
However this can be obtained with the element 2-2 in $M_d$ and then
with the same structure in both $M_d$ and $M_u$ \cite{dx}
\ini
M_d=\left( \begin{array}{ccc}
          0 & A & 0 \\
          A & D & B \\
          0 & B & C
          \end {array}\right),~
M_u=\left( \begin{array}{ccc}
          0 & A' & 0 \\
          A'^* & D' & B' \\
          0 & B'^* & C'
          \end {array}\right),
\fin
with a relation similar to (26), and
\ini
\frac{V_{ub}}{V_{cb}}\simeq \sqrt{\frac{m_u}{m_c}}.
\fin
In this case only the first generation gets mass by just mixing.
Models that lead to matrices similar to (31) have been studied by several authors
\cite{fx} (the phase in $B'$ can be shifted to $A$, or supposed to be zero).
A flavor permutation symmetry breaking \cite{psim} is often used.
In refs.\cite{chf} and \cite{nmf} the mass matrices are written as
\ini
M_d \simeq \left( \begin{array}{ccc}
          0 & \sqrt{m_d m_s} & 0 \\
          \sqrt{m_d m_s} & m_s & \sqrt{m_d m_b} \\
          0 & \sqrt{m_d m_b} & m_b
          \end {array}\right),
~M_u \simeq \left( \begin{array}{ccc}
          0 & \sqrt{m_u m_c} & 0 \\
          \sqrt{m_u m_c} & m_c & \sqrt{m_u m_t} \\
          0 & \sqrt{m_u m_t} & m_t
          \end {array}\right),
\fin
yielding a new successful relation
\ini
V_{cb} \simeq \sqrt{\frac{m_d}{m_b}}.
\fin
As hermitian matrices, such models have four texture zeros
(two zeros in symmetric positions are counted as one texture zero).
A systematic analysis of five texture zero matrices, in the $SO(10)$ model,
has been carried out by Ramond, Roberts, and Ross (RRR) \cite{rrr}.
They found five solutions which were consistent with low energy data
(all matrices have an approximate hierarchical expression in terms of
powers of $\lambda$;
for a matching of RRR analysis with the idea of naturalness see \cite{pw}).
Let us consider their solution 3:
\ini
M_d=\left( \begin{array}{ccc}
          0 & A & 0 \\
          A^* & B & C \\
          0 & C & D
          \end {array}\right),~
M_u=\left( \begin{array}{ccc}
          0 & 0 & C' \\
          0 & B' & 0 \\
          C' & 0 & D'
          \end {array}\right).
\fin
The phase in $A$ can be shifted to $C'$
and in ref.\cite{hw} and ref.\cite{chf} (see also \cite{gng}) these matrices are
written as
\ini
M_d \simeq \left( \begin{array}{ccc}
          0 & \sqrt{m_d m_s} & 0 \\
          \sqrt{m_d m_s} & m_s & \sqrt{m_d m_b} \\
          0 & \sqrt{m_d m_b} & m_b
          \end {array}\right),
~M_u \simeq \left( \begin{array}{ccc}
          0 & 0 & \sqrt{m_u m_t} \\
          0 & m_c & 0 \\
          \sqrt{m_u m_t} & 0 & m_t
          \end {array}\right),
\fin
leading to the mixings
\ini
V_{us} \simeq \sqrt \frac{m_d}{m_s},
\fin
\ini
V_{cb} \simeq \frac{m_s}{m_b},
\fin
coming from the down sector, while the mixing
\ini
V_{ub} \simeq \sqrt{\frac{m_u}{m_t}}
\fin
comes from the up sector.
Also consider the hierarchical expressions for eqns.(31) and (35):
\ini
M_d \simeq m_b \left( \begin{array}{ccc}
          0 & \lambda^3 & 0 \\
          \lambda^3 & \lambda^2 & \lambda^2 \\
          0 & \lambda^2 & 1
          \end {array}\right),~
M_u \simeq m_t \left( \begin{array}{ccc}
          0 & \lambda^6 & 0 \\
          \lambda^6 & \lambda^4 & \lambda^4 \\
          0 & \lambda^4 & 1
          \end {array}\right);
\fin
\ini
M_d \simeq m_b \left( \begin{array}{ccc}
          0 & \lambda^3 & 0 \\
          \lambda^3 & \lambda^2 & \lambda^2 \\
          0 & \lambda^2 & 1
          \end {array}\right),~
M_u \simeq m_t \left( \begin{array}{ccc}
          0 & 0 & \lambda^4 \\
          0 & \lambda^4 & 0 \\
          \lambda^4 & 0 & 1
          \end {array}\right).
\fin
Again the appearance of zeros and the hierarchical structure of mass matrices
point towards some broken horizontal symmetry
(for example $U(1)_H$ \cite{ir}) with a breaking
depending on the small parameter $\lambda$.
The texture zeros should be zero only up to the order that does not change
masses and mixings.
If the symmetry is exact,
only the third generation would be massive and the mixings vanish. Instead
the terms that break the symmetry, gradually fill in the mass matrices with
powers of $\lambda$, generating the hierarchy of masses and mixings. Hence
a broken symmetry can explain both the approximate zeros in mass matrices and
the hierarchy of non vanishing elements. Such a symmetry could also work in the
context of a unified or string theory.

Another approach to matrix models of fermion masses and mixings is the above
mentioned permutation symmetry breaking and a suggestive variation of it
\cite{usy}, which is called USY (Universal Strength of Yukawa couplings,
that is phase mass matrices).

To conclude this section we stress that two forms of quark mass matrices which
agree with numerical values of quark masses and mixing, namely eqn.(31) and
eqn.(35), are reported. The first has the same zeros in $M_u$ and $M_d$ and,
with a real $B'$ (only one phase is necessary for CP violation), contains
nine real parameters. The second has a non parallel structure of zeros and
contains eight real parameters. Of course, other forms may be considered
\cite{rbgg}. In the following section we use from the beginning the m.p.
basis approach to discover relations and properties of mass matrices.

\section{ Weak bases}

\noindent
In ref.\cite{blm} it was shown that, in the SM (with less than five
generations),
with a choice of the weak basis,
one can  always obtain the zeros of the Fritzsch model.
This is called the NNI (nearest neighbor interaction) basis.
Using transformations (4),(5) it is always possible to go to the NNI basis.
Then, using the rephasing of quark fields, one matrix has no phases and the
other two phases, twelve real parameters in all.
As the observables quantities are ten, it
is important to go to a basis where, after rephasing, ten parameters are left
(although not minimal the non hermitian NNI basis is interesting,
see ref.\cite{nni}).
Using transformations (4),(5) it is in fact
possible to get several interesting m.p. bases.

Let us begin with the case of only two generations. In such a case we have
five observables: four masses and one mixing angle. As shown in the
introduction, one can always diagonalize
$M_1$ and then use $V_2$ to obtain some special form of $M_2$. This is true
for an arbitrary number of generations,
allowing to shift the mixings to a single mass matrix.
Moreover, the diagonal matrix is real and
non negative. For example, if we set
$M_u=D_u$ and $M_d$ hermitian (and also real), then we get
(using the central values of quark masses, in GeV; $V \equiv V_{CKM}$)
\ini
M_d=V D_d V^+ = \left( \begin{array}{cc}
                   0.009 & 0.019 \\
                   0.019 & 0.088
                   \end {array}\right) \simeq            
m_s \left( \begin{array}{cc}
                   \lambda^2 & \lambda \\
                   \lambda & 1
                   \end {array}\right).
\fin            
Starting from $M_u=D_u$ and $M_d$ real and symmetric,
we can get a zero in $M_{d11}$ \cite{w} or $M_{d12}$ \cite{wz}, by means of a
rotation of right-handed fields.
Then $M_d M_d^+=V D_d^2 V^+$ gives
\ini
M_d=\left( \begin{array}{cc}
                   0 & a \\
                   b & c
                   \end {array}\right)=\left( \begin{array}{cc}
                                       0 & 0.021 \\
                                       0.023 & 0.088
                                     \end {array}\right),
\fin
with $M_{d12}\simeq M_{d21}$, and we recover the Weinberg model of section II,
or
\ini
M_d=\left( \begin{array}{cc}
                   a & 0 \\
                   c & b
                   \end {array}\right)=\left( \begin{array}{cc}
                                        0.021 & 0 \\
                                        0.088 & 0.023
                                      \end {array}\right),
\fin
with $M_{d11}\simeq M_{d22}$. Therefore, on these bases, the numerical values
of quark masses and mixing point towards a reduction of independent parameters
and to the relation (10).

For three generations again one can choose $M_u=D_u$ and $M_d$ hermitian
\cite{fkh,fk},
to yield (in the diagonal bases the non diagonal matrix has one physical phase
but three of them preserve arbitrary representation of $V_{CKM}$)
\ini
M_d=\left( \begin{array}{ccc}
          0.009 & 0.019 & 0.010 e^{i \varphi} \\
          0.019 & 0.093 & 0.113 \\
          0.010 e^{-i \varphi} & 0.113 & 2.995
          \end {array}\right)=
m_b \left( \begin{array}{ccc}
          0.003 & 0.006 & 0.003 e^{i \varphi} \\
          0.006 & 0.031 & 0.038 \\
          0.003 e^{-i \varphi} & 0.038 & 0.998
          \end {array}\right).
\fin
It is also possible to get a m.p. basis inside the NNI basis \cite{koide}.
Let us set $H=MM^+$. From arbitrary mass matrices first move to a basis with
$H_u=D_u^2$ \cite{bbhj}. Then, by means of a unitary matrix
\ini
U=\left( \begin{array}{ccc}
           1 & 0 & 0 \\
           0 & e^{i \alpha}c & e^{i \beta}s \\
           0 & -e^{i \gamma}s & e^{i \delta}c
          \end {array}\right),~\alpha+\gamma-\beta=\delta,
\fin
one can get $H'_u=UH_u U^+$, with $H'_{u12}=H'_{u13}=0$, and
$H'_d=UH_d U^+$, with $H'_{d12}=0$,
that is \cite{blm,koide} $M_d$ in the NNI form, and $M_u$ in the NNI form and
with the element 3-2 equal to zero.
Rephasing the quark fields only one phase remains in
$M_u$ and we have a m.p. basis.

Now we turn to m.p. bases with $M_u=D_u$ and $M_d$ not hermitian but
containing three zeros. In such cases equation
$M_d M_d^+=V D_d^2 V^+$ enable us to calculate $M_d$.
The case $M_d=D_d$ seems less interesting.
The first of such diagonal bases
to be studied \cite{fpr} has zeros in positions 1-1, 2-2 and 3-1.
This basis however does not seem to lead to useful relations.
The lower triangular form \cite{hs} gives the numerical values
\ini
M_d= \left( \begin{array}{ccc}
          0.023 & 0 & 0  \\
          0.104 & 0.106 e^{i \varphi} & 0 \\
          1.213 & 2.687 & 0.541
         \end{array} \right)
\fin
showing $M_{d21}\simeq M_{d22}$ and $M_{d32}\simeq 2 M_{d31}$. In ref.\cite{ft}
$M_d$ has the zeros in symmetric positions \cite{ma} and, with the updated
values of mixings,
\ini
M_d= \left( \begin{array}{ccc}
          0 & 0.023 & 0  \\
          0.021 & 0.104 e^{i \varphi} & 0.104 \\
          0 & 1.213 & 2.741
         \end{array} \right).
\fin
This suggests to take
\ini
M_d \simeq \left( \begin{array}{ccc}
              0 & a & 0 \\
              a & be^{i \varphi} & b \\
              0 & c & 2c
         \end{array} \right)
\simeq   \left( \begin{array}{ccc}
              0 & \sqrt{m_d m_s} & 0 \\
              \sqrt{m_d m_s} & m_s e^{i \varphi} & m_s \\
              0 & m_b/\sqrt{5} & 2m_b/\sqrt{5}
         \end{array} \right),
\fin
with the same 1-2 submatrix as in eqn.(20),
yielding the relations (10),
\ini
V_{cb} \simeq \frac{3}{\sqrt{5}}\frac{m_s}{m_b},
\fin
\ini
\frac{V_{ub}}{V_{cb}} \simeq \frac{1}{3} V_{us},
\fin
and the hierarchical expression
\ini
M_d \simeq m_b \left( \begin{array}{ccc}
              0 & \lambda^3 /\sqrt{2} & 0 \\
              \lambda^3 /\sqrt{2} & \lambda^2 /\sqrt{2} & \lambda^2 /\sqrt{2} \\
              0 & 1/\sqrt{5} & 2/\sqrt{5}
         \end{array} \right).
\fin

\begin{center}
\begin{tabular}{|c|cccccccccccc|}
\hline
$\delta_{CP}(^o)$ & 30 & 40 & 50 & 60 & 70 & 80 & 90 & 100 & 110 & 120 & 130 &
140 \\ \hline \hline
$|M_{d22}|$ & 124 & 121 & 117 & 112 & 107 & 100 & 94 & 86 & 79 & 70
& 62 & 54 \\ \hline
$M_{d23}$ & 78 & 83 & 88 & 94 & 100 & 107 & 113 & 119 & 124 & 129
& 132 & 136 \\ \hline
$-\varphi(^o)$ & 32 & 42 & 50 & 58 & 65 & 72 & 78 & 85 & 91 & 98 & 106 &
115 \\ \hline
\end{tabular}
\end{center}

$~$
\centerline{Table 1}
 
$~$

\noindent
For the Jarlskog's parameter $J$ \cite{jar}, which is related to CP violation
in the SM and given by the relation
\ini
$det$ [H_d,H_u]=2B \cdot T \cdot J,
\fin
where $B=(m_b^2-m_d^2)(m_b^2-m_s^2)(m_s^2-m_d^2)$,
$T=(m_t^2-m_u^2)(m_t^2-m_c^2)(m_c^2-m_u^2)$, and
$J=(-1)^{r+s} $Im$ (V_{ij}V_{kl}V^*_{kj}V^*_{il})$ is obtained crossing out
row $r$ and column $s$ of $V_{CKM}$,
we have the approximate expression
\ini
J \simeq \frac{3}{5} \frac{m_d m_s}{m_b^2} \sin \delta_{CP}.
\fin
It can be checked that the relation between $M_{d22}$ and $M_{d23}$ is sensitive
to variations of the CP violating phase $\delta_{CP}$ in $V_{CKM}$.
Equality $M_{d23}=|M_{d22}|$ is obtained for $\delta_{CP}=75^o$,
which is very close
to the experimentally favoured value \cite{pdg,cp}. For $\delta_{CP}=126^o$,
which is near the bound of the allowed region, it happens $M_{d23}=2 |M_{d22}|$.
All others parameters in $M_d$ are nearly independent from $\delta_{CP}$.
In table 1 we report the values of $M_{d22}$, $M_{d23}$ (in MeV) and $\varphi$
versus $\delta_{CP}$. Note also that for $\delta_{CP}=90^o$ it is
$|M_{d22}|=m_s$. 

One can worry about the relation $M_{d33} \approx 2 M_{d23}$,
but this depends on the value of $V_{ub}$, which has large uncertainties;
for example, taking $V_{ub}=0.0035$, the above relation is well satisfied.
Therefore, the study of this basis, which has $M_u$ diagonal and three zeros
built in $M_d$, leads to three simple relations between matrix elements
of $M_d$. With the actual hierarchical values of quark masses and mixings,
these relations are related to the value of the phase $\delta_{CP}$, to the
relation (10), and to the value of $V_{ub}$, respectively.

This last diagonal three zero basis can be obtained from the first by a unitary
transformation of the form (46) on $d_R$.
Relabeling indices for the three diagonal bases leads to other fifteen
bases, that however do not seem more interesting than the basis (48).
Actually, starting from the first of such bases, other thirty-five bases
with $M_u$ diagonal and $M_d$ with three zeros can be obtained by means of
$U(2)$ transformations similar to (46) on $d_R$ and relabeling.
In the same way
other seventeen bases can be obtained from the triangular basis.  
One of these is studied in detail in ref.\cite{kuo}.

As concerning hermitian non diagonal bases we refer to the paper \cite{bbhl},
where eighteen bases
without zeros on the diagonal are obtained. Starting from $M_1$ diagonal and
$M_2$ hermitian, by the unitary transformation (46)
one yields $U M_1 U^+=M_1'$
with texture zeros in positions 1-2 and 1-3, and $U M_2 U^+=M_2'$
with a texture zero in position 1-2 or 2-3. Then by relabeling indices
other seventeen bases
are obtained. Looking at the five RRR solutions, one realizes that solutions
1 \cite{h}, and  3, 4, 5, correspond to four of such eighteen bases
with the element 1-1 of both mass matrices set equal to zero,
and $M_{d23} \simeq 2 M_{d22}$.
In particular, matrices (36)
have $M_{11}=0$ and $M_{d23} \simeq M_{d22}$.
On the contrary, matrices (31) cannot be obtained from a m.p. basis, because
they already contain ten parameters (eight moduli and two phases), although
one can always take $M_{13}=0$ in both hermitian matrices \cite{fri}.

\section{ Concluding remarks}

\noindent
In summary, some of the m.p. bases clearly show relations among the matrix
elements and/or vanishing elements.
In the SM different m.p. bases with different basis zeros
lead to the same values of the ten observables. Instead, in left-right symmetric
models, different zeros correspond to different values of the
(more than ten) observables.
Actually, in such models, one can often get one diagonal matrix, but the other
has no freedom, because $V_u=V_d$,
and then putting zeros corresponds to choosing ansatze.
These must give the same left-handed currents but can produce different
right-handed currents, which have not been observed till now, because the
corresponding gauge bosons are expected to be heavy, but play an important role
in unified theories, and affect proton decay \cite{ab}.
For example, let us look at matrix (48). Viewed in the SM, such a matrix shows
a hierarchical non symmetric structure with three simple relations between
elements, while the zeros have no physical content.
Instead, in a left-right model also the zeros (which in this case could be
approximate) have a physical meaning, and the implications of matrix (48)
are quite different, for example, from those of matrix (47).
In a similar way, in left-right models, it is always possible to take a basis
where both $M_d$ and $M_u$ have the zeros as in eqn.(31) \cite{blm},
while taking one matrix diagonal, or the symmetric form (31),
or the NNI form, gives different definite predictions.
It is still left for the future to confirm the need for an extension of
the gauge group at higher energy and,
in such a case, the selection of the right fermion mass
matrix forms.

In this paper we have considered quark mass matrices. Of course, a similar
work can be done for charged lepton and neutrino mass matrices, although we do
not know neutrino masses so well. In unified theories lepton mass matrices
are related to quark mass matrices, but other analogies can exist between
the quark and the lepton sector. For example, in ref.\cite{nmf} all fermion
mass matrices have the same texture zeros,
and in ref.\cite{fal} a strict similarity in the Dirac sector is assumed;
see also ref.\cite{kaki} for a suggestive approach based on permutation
symmetry. Neutrino masses and mixings
are in the main stream of current physical research.

$~$

I thank prof. F. Buccella for critical comments. I thank also
F. Tramontano for useful discussions, and Carlo Nitsch for a comment
on weak bases.

\end{document}